\documentclass[fleqn,usenatbib]{mnras}

\usepackage{newtxtext,newtxmath}

\usepackage[T1]{fontenc}

\DeclareRobustCommand{\VAN}[3]{#2}
\let\VANthebibliography\thebibliography
\def\thebibliography{\DeclareRobustCommand{\VAN}[3]{##3}\VANthebibliography}

\usepackage{graphicx}	
\usepackage{amsmath}	

\title[Fast radio bursts trigger aftershocks]{Fast radio bursts trigger aftershocks resembling earthquakes, but not solar flares }

\author[T. Totani \& Y. Tsuzuki]{
Tomonori Totani$^{1,2}$\thanks{E-mail: totani@astron.s.u-tokyo.ac.jp}
and Yuya Tsuzuki$^{1}$
\\
$^{1}$Department of Astronomy, School of Science, The University of Tokyo, Bunkyo-ku, 
Tokyo 113-0033, Japan \\
$^{2}$Research Center for the Early Universe, School of Science, The University of Tokyo, 
Bunkyo-ku, Tokyo 113-0033, Japan}

\date{Accepted XXX. Received YYY; in original form ZZZ}

\pubyear{2023}

\begin{document}
\label{firstpage}
\pagerange{\pageref{firstpage}--\pageref{lastpage}}
\maketitle

\begin{abstract}
The production mechanism of repeating fast radio bursts (FRBs)
is still a mystery, and correlations between burst occurrence
times and energies may provide important clues to elucidate it. 
While time correlation studies of FRBs have been mainly performed using wait time distributions, 
here we report the results of a correlation function analysis of repeating FRBs in the 
two-dimensional space of time and energy. 
We analyze nearly 7,000 bursts reported 
in the literature for the three most active sources of FRB 20121102A, 20201124A, and 20220912A,
and find the following characteristics that are universal in the three sources. 
A clear power-law signal of the correlation function is seen, extending to the typical burst 
duration ($\sim$ 10 msec) toward shorter time intervals ($\Delta t)$. 
The correlation function indicates that every single burst has about a 
10--60\% chance of producing an aftershock at a rate decaying 
by a power-law as $\propto (\Delta t)^{-p}$ with $p =$ 1.5--2.5,
like the Omori-Utsu law of earthquakes.
The correlated aftershock rate is stable regardless of source activity changes, 
and there is no correlation between emitted energy and $\Delta t$.
We demonstrate that all these properties are quantitatively common to 
earthquakes, but different from solar flares in many aspects,  
by applying the same analysis method for the data on these phenomena. 
These results suggest that repeater FRBs are a phenomenon in which energy stored in rigid 
neutron star crusts is released by seismic activity.
This may provide a new opportunity for future studies to explore the physical properties 
of the neutron star crust.
\end{abstract}




\section{Introduction}

Fast radio bursts (FRBs) are extragalactic transient phenomena that shine 
in radio wavelengths for short durations lasting only 1--10 milliseconds
\citep[see][for reviews]{Cordes2019,Platts2019,Zhang2020,Petroff2022}. 
Some FRB sources are known to produce many bursts repeatedly. 
Repeaters are thought to be neutron stars, but the causes of bursts and the radiation 
mechanism are not well understood. 

More than several thousand bursts have already been detected from a few repeaters,
and a detailed statistical analysis of their occurrence times and energies may provide some 
information on the burst production mechanism.
Previous studies on time correlations of repeating FRBs 
have been conducted using the distribution of wait times between two successive bursts
\citep{Wang2017,Oppermann2018,Wang2018,Zhang2018,Li2019,Gourdji2019,Wadiasingh2019,Oostrum2020,Tabor2020,Aggarwal2021,Cruces2021,Li2021,Zhang2021,Hewitt2022,Xu2022,Zhang2022,Du2023,Jahns2023,Sang2023,Wang2023,Zhang2023,Zhang2023b}.
The wait time distribution is known to be bimodal with a boundary of about 1 second, 
and the distribution on the long time side can be described by an uncorrelated 
Poisson process \citep{Cruces2021,Hewitt2022,Jahns2023}.
The short-side distribution is thought to reflect a time scale related 
to the physical activity of the source or radiative processes, but its origin is unknown.

The wait time distribution does not take full advantage of time-related statistics,
because time correlations may exist not only between consecutive bursts but also 
across other bursts. Therefore in this study, we directly calculate the two-point
correlation function in two-dimensional
space of occurrence time $t$ and emission energy $E$ of FRBs, using the method 
widely used in cosmology to study large-scale structures. 
We collected a wide range of reported observations 
including a large number of repeating FRB events from the three most active sources of
FRB 20121102A, 20201124A, and 20220912A
\citep{Li2021,Hewitt2022,Xu2022,Zhang2022,Jahns2023,Zhang2023}, 
and nearly 7,000 bursts in total are analyzed to compute the correlation function. 

At least some FRBs are known to occur in magnetars (neutron stars with extremely strong 
magnetic fields of $\gtrsim 10^{14}$ G) \citep{Bochenek2020,CHIME2020-Galactic}, 
and explosive
phenomena in magnetars are believed to be triggered by starquakes in neutron star crusts, 
which is induced by magnetic energy \citep{Kaspi2017}.
For this reason, various phenomena of neutron stars have been compared to 
earthquakes and solar flares, and similarities are often discussed \citep{Cheng1996,Gogus1999}.
Therefore we will use the same method to analyze the time-energy correlation function of
earthquake and solar flare data,
and examine the similarities with the statistical properties of repeating FRBs.

\begin{table*}
\footnotesize
\centering
\caption{Summary of the FRB data sets}
\begin{tabular}{lcccccccccc}
\hline
Data set name &Telescope& Period & Days$^a$ & ${t_{\rm obs}}^b$ & Events & ${r_m}^c$
 & ${C_{\mathrm{best}}}^d$ & $p_{\mathrm{best}}$&$\tau_{\mathrm{best}}$&  $n^e$\\
 && (MJD) &  & (day) & & (day$^{-1}$) &$C_{-1\sigma}$& 
 $p_{-1\sigma}$&$\tau_{-1\sigma}$& \\
 & &  &  & & &&$C_{+1\sigma}$& $p_{+1\sigma}$&$\tau_{+1\sigma}$& \\
\hline
FRB 20121102A (L21) &FAST& 58724.87--58776.88 & 39 & 1.76 & 1651
         &1500&5100&1.6&0.0020&0.28 \\
  &&&&&&&3100&1.4&0.0009&\\
  &&&&&&&9700&1.8&0.0033&\\
 \hline
FRB 20121102A (H22) &Arecibo& 57510.80--57666.42 & 18 & 0.733 & 475
         &870 & 490 & 9.1 & 0.28 &0.17\\
 &&&&&&& 280 & 2.1 & 0.019 &\\
 &&&&&&& 1200 & $> 30$ & $>$1.4 &\\
 \hline
FRB 20121102A (J23)  &Arecibo& 58409.35--58450.28 &8 &0.272 &1027 
                &4900 & 770 & 2.3 & 0.012 & 0.40\\
 &&&&& (849)$^f$ && 500 & 1.8 & 0.0063\\
 &&&&&&& 1100 & 3.5 & 0.028 &\\
 \hline
FRB 20201124A (X22) &FAST& 59307.33--59360.18 &45 &3.13 &1863 
         &840& 340 & 28.3 & 1.3 & 0.16\\
  &&&&&&& 250 & 4.5 & 0.13 &\\
  &&&&&&& 500 & $>30$ & $>$1.5 &\\
 \hline
FRB 20201124A (Z22 D3) & FAST & 59484.81--59484.86 &1&0.040&232
         &5800& 270 & 3.4 & 0.071 & 0.54 \\
 &&&&&&& 83 & 1.5 & 0 \\ 
 &&&&&&& $\infty$ & $>30$ & $>$1.7 &\\
 \hline
FRB 20201124A (Z22 D4) & FAST & 59485.78--59485.83 &1&0.040&542
         &14000& 54 & 4.2 & 0.19 & 0.50 \\
 &&&&&&& 35 & 2.1 & 0.058 &\\
 &&&&&&& 60 & $>30$ & $>$1.9 &\\
 \hline
FRB 20220912A (Z23) & FAST & 59880.49--59935.39 & 17 & 0.32 & 1076
         & 6900 & 70 & 5.7 & 0.26 & 0.30 \\
 &&&&&&& 50 &  2.4 & 0.043 &\\
 &&&&&&& 170 & $>30$ & $>$1.8 &\\
\hline
\end{tabular}
\raggedright \\
$^a$Total number of days on which observations were made \\
$^b$Total observation time \\
$^c$Mean event rate weighted by the number of pairs for all observation dates \\
$^d$The best-fit value, 68\% confidence level (CL) lower and upper limits of the parameters 
in the fitting of $\xi(\Delta t) = C (\Delta t + \tau)^{-p} / \tau^{-p}$,
where $\tau$ in [s] \\
$^e$The branching ratio $n = \int r_m \, \xi(\Delta t) d(\Delta t)$ \\
$^f$When only independent events are counted by grouping sub-bursts together
\label{table:FRB}
\end{table*}

\section{Correlation function study on FRBs}

\subsection{The FRB data sets}

We compute correlation functions for the seven data sets (Table \ref{table:FRB}) of FRBs
observed by the two radio telescopes (Arecibo and FAST)
from the following three FRB repeaters.

FRB 20121102A is the first discovered \citep{Spitler2014,Spitler2016}, 
highly active, and most well-studied repeater
located in a star-forming dwarf galaxy at redshift 
$z = 0.193$ \citep{Bassa2017,Chatterjee2017,Tendulkar2017}. 
The solar system barycentric time $t$
and emitted energy $E$ of bursts were taken from tables given in
the original papers of the L21 \citep{Li2021}, H22 \citep{Hewitt2022}, and 
J23 \citep{Jahns2023} data sets. The observation logs 
(start and end times of each observation) are given in the papers of H22 and J23,
and we obtained that for the L21 data directly from the authors. 
In the J23 data, all 1,027 sub-bursts in the 849 independent events are listed 
in their table, and we used all sub-bursts in the main analysis.

FRB 20201124A \citep{CHIME2021} is another repeater known for its high activity.
It is in a Milky Way-like, barred spiral galaxy
at $z=0.0979$ \citep{Xu2022}. Barycentric times of the X22 \citep{Xu2022} and 
Z22 \citep{Zhang2022} data are
from the tables in the original papers. Burst energies of X22 were calculated
from fluence and signal bandwidth given in their table, using
$z$ and the corresponding luminosity distance, while for Z22 we used ``the energies calculated
with the central frequency'' given in their table. We obtained the observation logs
of the X22 data set directly from the authors. 
This repeater was extremely active in the Z22 data, with the highest event rate detected 
from a single FRB source. Particularly large numbers of bursts were detected 
on days three and four, and therefore,
we analyzed the third and fourth days as independent data sets 
(Z22 D3 and Z22 D4, respectively) to examine the
dependence on activity level.

FRB 20220912A \citep{McKinven2022} is located at
a position that is consistent with a likely host galaxy of stellar mass 
$\sim 10^{10} M_\odot$ at  $z=0.0771$ \citep{Ravi2022}.
We took barycentric times and energies of the 1,076 bursts 
as the Z23 \citep{Zhang2023} data from the table in the original paper. 

The redshifts of these FRB sources are not large enough for cosmological effects 
to be significant, and hence no correction for the cosmological time dilation is made.

\subsection{Correlation function calculations}

\begin{figure*}
\includegraphics[width=12cm,angle=-90]{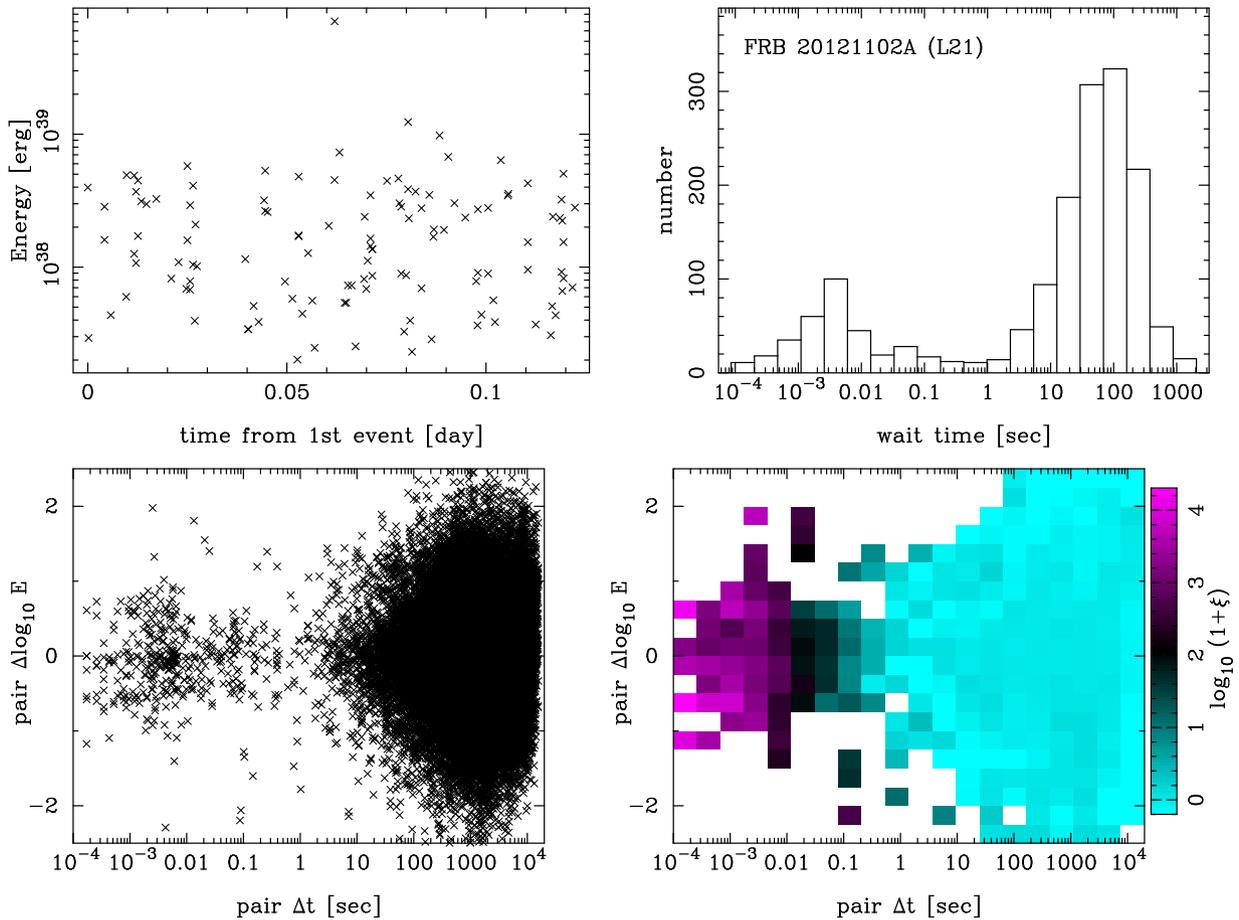}
\caption{Time-energy space correlation analysis for the FRB 20121102A L21 data. 
The $t$-$E$ plot of events on the day with the highest event rate
(upper-left), the wait time distribution (upper-right),
the $\Delta t$-$\Delta \lg E$ plot of pairs (lower-left), and the correlation function
in $\Delta t$-$\Delta \lg E$ space (lower-right) are shown.
}
\label{fig:FRB_L21}
\end{figure*}

For the difference of time ($\Delta t \equiv t_2 - t_1$, $t_2 > t_1$) and 
energy ($\Delta \lg E \equiv \lg E_2 - \lg E_1$, where $\lg = \log_{10}$) of
a burst pair (burst 1 and 2), the two-point correlation function $\xi(\Delta t, \Delta \lg E)$ 
is defined as the excess of the number of pairs $N_p$ over the uncorrelated case:
\begin{equation}
dN_p = (1 + \xi) \, \bar n_p \, d(\Delta t) \, d(\Delta \lg E) \ , 
\end{equation}
where $\bar n_p$ is the expected pair number density in the uncorrelated case.

The method of estimating the correlation function is widely used in cosmology 
to study the spatial correlation of galaxies, and has been described in many references
[e.g. \S16.7 of \citet{Peacock1999}]. 
The two-point correlation function is estimated by counting the number of pairs consisting 
of two objects and comparing them to a random sample made up of uncorrelated objects. 
To reduce statistical error, random data are usually generated with a much larger number of
objects than the observed data, and in this study, the random sample is 100 times larger 
($10^4$ times larger in terms of the number of pairs).
The number of all possible pairs involved in a given bin of $\Delta t$-$\Delta \lg E$ 
space is denoted as $DD$ for the real data sample and $RR$ for the random sample,
where $DD$ and $RR$ are appropriately normalized for different sample sizes. 
Then the correlation function can be estimated as
\begin{equation}
\xi(\Delta t, \Delta \lg E) = \frac{DD}{RR} - 1 \ , 
\end{equation}
but here we use
\begin{equation}
\xi(\Delta t, \Delta \lg E) = \frac{DD - 2DR + RR}{RR}  \ , 
\end{equation}
which is an estimator that has less variance and is most widely used \citep{Landy1993},
where $DR$ is the number of cross-pairs between the data and random samples. 
(We confirmed that using the former ``natural''
estimator has little effect on the conclusions of this paper.)
The time correlation function $\xi(\Delta t)$ is estimated in the same way,
but without binning to the energy direction. 

The random data to calculate $\bar n_p$ 
were generated as follows. The observation period for a single FRB data 
set spans multiple days, but a continuous observation is only a few hours per day.
Therefore, we assume that it is a Poisson process with a constant rate and 
a constant energy distribution within a day. There are occasional interruptions
during one day of observation, and these were taken into account in the random data generation
according to the observation logs.
The energy $E$ of the random data was generated assuming a 
cumulative distribution $f({<}E)$, which
was empirically constructed from the observed energies as
$f({<}E_i) = i/(1+N)$ ($i = 0, \dots , N+1$), 
where $f({<}E)$ is linearly interpolated between $E_i < E < E_{i+1}$. Here, $E_i$ 
are the energies (in increasing order) of the observed $N$ event in one day
for $i = 1, \dots , N$, and we set
 $E_0 = E_1 - \delta E$ and $E_{N+1} = E_N + \delta E$,
where $\delta E = (E_N - E_1) / (N-1)$ is the mean energy separation. 
Pairs across different days were not considered. Then pair counts within a day were 
added up for all observed days to compute the correlation 
function for a single data set.

The simplest way to estimate the correlation function errors
is by Poisson error for the pair counts. However, if the correlation is nonzero, 
the Poisson error is a lower bound to the real error, and furthermore, 
errors between different bins are correlated. 
We, therefore, computed the covariance matrix $C_{ij}$ of $\xi(\Delta t)$
using the jackknife method \citep{Norberg2009} 
by dividing the observation time of each day into 20 segments, 
where $i$ and $j$ denote the bins of $\Delta t$. 
The error $C_{ii}^{1/2}$ for the $i$-th $\Delta t$ bin determined by this method 
is not significantly different from that evaluated by Poisson statistics.
Jackknife errors are sometimes smaller than Poisson errors, which is probably due to
a sampling bias. Therefore, the larger of Poisson and jackknife errors are conservatively shown
at each data point when errors of $\xi(\Delta t)$ are presented.

The catalog of time and energy obtained from observations is subject to bias 
due to selection effects. The most obvious effect is that bursts of weak radio fluxes are 
below the detection limit and do not enter the sample. 
However, since we are empirically constructing the energy distribution function for the
random samples from the observed data, 
the incompleteness with respect to energy is correctly taken into account here.
Another important effect is that when two independent events occur in close proximity in time, 
they are indistinguishable from sub-bursts within a single event, or the detection sensitivity 
to fainter bursts is reduced. This effect is, in fact, important in the FRB samples analyzed here, 
and it will be discussed in detail in the next section.

\subsection{Results}

\begin{figure}
\includegraphics[width=13cm,angle=-90]{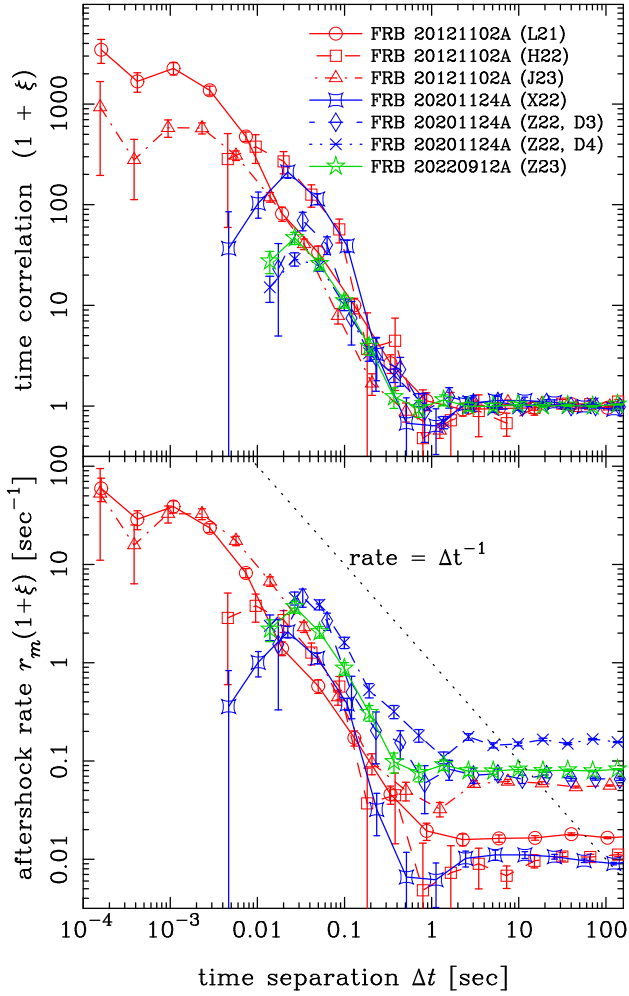}
\caption{ The time correlation function $\xi(\Delta t)$ of the seven FRB data sets.  
The upper panel shows $(1+\xi)$, while the lower panel shows $r_m (1+\xi)$, the
aftershock rate occurring $\Delta t$ after an event, where $r_m$ is
the mean event rate.  }
\label{fig:FRB_xi1D}
\end{figure}

\begin{figure*}
\includegraphics[width=7cm,angle=+90]{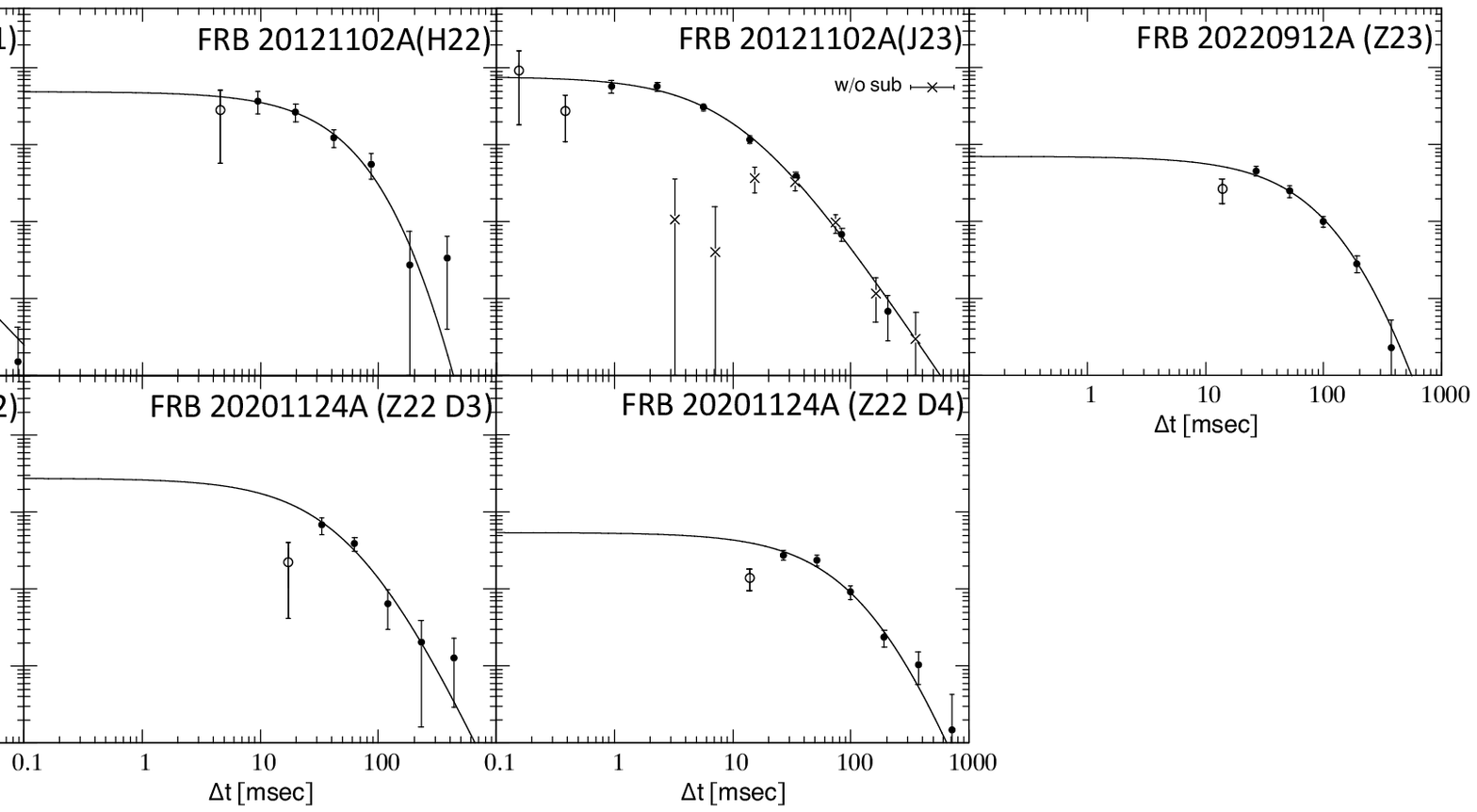}
\caption{ Model fits to the time correlation functions of FRBs. 
To avoid the effects of removing sub-bursts, open-circle data points were not used for fitting.
For the FRB 20121102A (J23)
data, two cases are plotted: one in which all sub-bursts are considered
(filled and open circles, the fit is to this case), and the other in which only the first sub-burst of
a single independent event is considered (crosses).}
\label{fig:FRB_fits}
\end{figure*}

Figure \ref{fig:FRB_L21}
shows the event distribution in $t$-$E$ space, the wait time distribution, 
and the pair distribution and correlation function $\xi$ in $\Delta t$-$\Delta \lg E$ space
for the FRB 20121102A L21 data set (the other six are shown in Figs. 
\ref{fig:FRB_H22}--\ref{fig:FRB_Z23} in Appendix \ref{app:figs}). 

The time correlation functions $\xi(\Delta t)$ 
for the seven data sets are shown together in Fig. \ref{fig:FRB_xi1D} 
and separately in Fig. \ref{fig:FRB_fits}. 
No clear signal in the $\Delta t > 1$ s 
region indicates that the uncorrelated Poisson process can describe the data, which confirms previous studies. On the other hand, 
clear correlation signals are detected for all the data sets at $\Delta t \lesssim 1$ s,
corresponding to the shorter side of the bimodal wait time distribution. 
These signals exhibit power-law behavior in $\Delta t \sim$ 0.01--1 s, 
but are flatter in the shortest $\Delta t$ region. In the region of $\Delta t \lesssim$ 0.01 s, 
the signal behavior differs among the seven data sets, likely due to different treatments 
about sub-bursts in a single event. In general, if the radio flux rises again before falling 
to noise levels, it is considered a sub-structure within a single event. 
However, the exact criteria and whether to include sub-bursts in the catalog are not 
standardized \citep{Hewitt2022,Jahns2023}. This effect is clearly seen in the 
FRB 20121102A (J23) data set, 
where every sub-burst is explicitly listed in the catalog. 
When only the first sub-burst of an independent event (849 in total) is used in the analysis,  $\xi$ 
becomes flat below 30 msec, whereas $\xi$ continues to increase 
smoothly with decreasing $\Delta t$ until 1 ms when all 1,027 sub-bursts are used
(Fig. \ref{fig:FRB_fits}). 
This suggests that even those classified as sub-bursts might be better treated as 
independent events.

We fit these signals with the function 
\begin{equation}
\xi(\Delta t) = C \, \frac{(\Delta t + \tau)^{-p}}{\tau^{-p}} \ , 
\label{eq:fit_function}
\end{equation}
by minimizing $\chi^2$. Ideally, $\chi^2$ should be calculated using the inverse of 
the covariance matrix \citep{Norberg2009}, $C^{-1}_{ij}$, 
but it is known that when the sample size is limited, inverting $C_{ij}$ 
estimated by jackknife can be numerically unstable \citep{Okumura2008,Pope2008}. 
In fact, we found that $\chi^2$ using $C_{ij}^{-1}$ 
varies sensitively with changes in the analysis parameters
(e.g. binning), and reliable results could not be obtained. 
Since a precise determination of the model parameter values is not the main purpose of this study,
error correlations between different $\Delta t$ bins were not considered, and 
the larger of the Poisson and jackknife errors were employed for the error
of the $i$-th $\Delta t$ bin. Perhaps the best way to properly incorporate error correlation
is to perform Monte Carlo simulations 
with realistic theoretical modeling, which is beyond the scope of this study.
The upper and lower 1$\sigma$ (68\% CL) limits for the three model parameters 
($C$, $p$, and $\tau$) were determined by
the excess $\Delta \chi^2$ from the minimum, which is expected to
follow a $\chi^2$ distribution with three degrees of freedom.

As seen in the case of FRB 20121102A (J23) (Fig. \ref{fig:FRB_fits}), the artificial
effect of sub-burst removal results in a sharp drop of $\xi(\Delta t)$ in
the smallest $\Delta t$ region. The drops seen in other data sets are also likely due to this effect,
and fitting the functional form of eq. (\ref{eq:fit_function}) to such a data set may 
produce biased results. Therefore, we excluded data points at $\Delta t$ smaller than 
the peak of $\xi(\Delta t)$ from the fit. 
In the large $\Delta t$ region ($\gtrsim$ 1 s), no significant correlation signals are found.
However, because of the large number of available pairs in this region, the statistical errors are
very small, and consequently the fit becomes sensitive to systematic uncertainties. 
In particular, our analysis assumes that the burst rate is constant within a day, but 
if the actual rate is slowly changing on a time scale of about a day, 
it may appear as a non-zero signal in the correlation function.
We are interested in the $\xi$ signal clearly detected at $\Delta t < 1$ s, and since
it decreases rapidly as $\Delta t$ increases, 
the signal is hidden by noise in large $\Delta t$ regions.
Therefore, data at $\Delta t$ larger than the point at which $\xi$
first becomes negative were excluded from the fit. 

The best-fit parameter values and their errors of $C$, $p$, and $\tau$ are
shown in Table \ref{table:FRB}, and the best-fit curves are
shown in Fig. \ref{fig:FRB_fits}. 
The power law fits are particularly good for the two data sets (FRB 20121102A
L21 and J23) extending to
small $\Delta t$ comparable with typical FRB durations (1--10 msec), 
with the index $p \sim$ 1.6 or 2.3. 
The fact that $p$ is restricted to a relatively narrow range indicates that these data
cannot be fitted by other function forms, such as the exponential function.
For the FRB 20201124A (Z22 D3) data, pure power-law without the effect of 
non-zero $\tau$ is also consistent with the data within 68\% CL, 
and no lower bound on $\tau$ (and corresponding upper bound 
on $C$) can be placed. 

The data sets other than FRB 20121102A (L21) and (J23)
are likely biased toward larger $\tau$ because small $\Delta t$ pairs have been removed 
as sub-bursts, thus making the power-law features less clear. However,
the power-law with $p\sim 2$ is within the 1$\sigma$ error range for all FRB data sets except 
for FRB 20201124A (X22).
The $p$ value of the X22 data may be significantly larger than those in other sets, but 
it is due to the only data point with a large error 
at $\Delta t = $ 230 msec (Fig. \ref{fig:FRB_fits}), and hence
the possibility of statistical fluctuation cannot be ruled out. The lower limits of 2 and 3$\sigma$ 
are $p=$ 3.0 and 2.1, respectively, for the X22 data.  It should also be noted that
$p$ tends to be larger when $\tau$ is large due to degeneracy in the fit
(see the next paragraph), and
both $\tau$ and $p$ may be biased to a greater value because of sub-burst removal. 

As for the upper limits on $p$, 
the data sets other than FRB 20121102A (L21) and (J23) are consistent with
extremely large values of $p > 30$ within $1 \sigma$. This means that $\xi(\Delta t)$ 
can also be fitted with an exponential function, because the fitting function
converges into $\exp( - \Delta t / \tau')$ in the limit of $\tau \rightarrow \infty$
and $p \rightarrow \infty$ with $\tau' = \tau / p$, which can be shown as
\begin{equation}
\frac{\xi(\Delta t)}{C} = \left( 1 + \frac{\Delta t}{\tau} \right)^{-p} = 
\left( 1 + \frac{1}{\alpha} \right)^{-\alpha (\Delta t / \tau) p}
\rightarrow \exp \left( - \frac{\Delta t}{\tau'} \right) \ ,
\end{equation}
where $\alpha \equiv \tau / \Delta t$.
Our parameter search is limited to $p \leq 30$, and if $p=30$ falls within 
the 1$\sigma$ error region, ``$p > 30$'' and the 
corresponding value of $\tau$ are indicated in 
Table \ref{table:FRB} as 1$\sigma$ upper bounds, but this is essentially
an exponential fit with $\exp[-\Delta t / (\tau / p) ]$.

In summary, at least two of the data sets show clear power laws
of $p\sim 2$, and the other data sets are consistent with similar power laws. 
The power law feature indicates that there is no characteristic time scale for the correlation.
It is then natural to assume that the shorter peak of the bimodal wait-time distribution
does not reflect the activity duration, but rather correlated aftershocks like earthquakes.
In fact, it is known that the aftershock occurrence rate of earthquakes follows a power-law
$\propto (\Delta t + \tau)^{-p}$ with $p \sim 1$ 
(the Omori-Utsu law) \citep{Omori1894,Utsu1957,Utsu1961,Utsu1995,deArcangelis2016}.
This consideration provides a good motivation to investigate the similarity of 
correlation functions between FRBs and earthquakes.

\section{Comparison with earthquakes and solar flares}

\subsection{The earthquake data sets}

\begin{table*}
\footnotesize
\centering
\caption{Summary of the earthquake data sets}
\begin{tabular}{cccccccccccc}
\hline
Name & Longitude & Latitude & Period & ${t_{\rm obs}}^a$ & Events & 
${r_m}^b$ & ${C_{\mathrm{best}}}^c$ & 
$p_{\mathrm{best}}$&$\tau_{\mathrm{best}}$&  $n^d$\\
 & (deg) & (deg) &   (MJD) & (day) & &(day$^{-1}$)&$C_{-1\sigma}$& $p_{-1\sigma}$
  &$\tau_{-1\sigma}$& \\
 &&&   & & &&$C_{+1\sigma}$& $p_{+1\sigma}$&$\tau_{+1\sigma}$& \\
\hline
Narita b311$^e$ & 140.10--140.70 & 35.70--36.10 & 52500--53500 & 1000 & 938        
        &1.1& 42 & 0.88 & 112 & 0.31\\
 &&&& &&& 22 & 0.73 & 20 & 0.60 \\
 &&&& &&& 110 & 1.1 & 460 &\\
 \hline
Narita a311$^e$ & 140.10--140.70 & 35.70--36.10 & 55700--55900 & 200 & 997
          &5.2& 5.8 & 1.1 & 270 & 0.15\\
 &&&  &&&& 3.2 & 0.75 & 80 & 0.32 \\
 &&&  &&&& 12 & 2.3 & 1800 &\\
 \hline
 Choshi & 140.81--140.96 & 35.74--35.86 & 52500--55600 & 3100 &800
                 &0.35& 130 & 1.0 & 180 & 0.37 \\
&&&&&       && 62 & 0.87 & 20 & 0.51 \\
&&&&&       && 370 & 1.37 & 740 &\\
 \hline
 Kanto & 139.30--141.10 & 35.00--36.00 & 53000--53500 & 500 &1399 
         &2.9& 34 & 0.75 & 16 & 0.17 \\
 &&&&  &&& 16 & 0.7 & 2.0 & 0.44 \\
 &&&&  &&& 152 & 0.87 & 64 &\\
 \hline
Izumo & 132.20--132.60 & 34.70--35.50 & 52500--55500 & 3000 &1596
        &0.57& 380 & 0.82 & 5.0 & 0.18 \\
 &&&&  &&& 140 & 0.72 & 0.6 & 0.34 \\
 &&&&  &&& 1800 & 0.98 & 24 &\\
\hline
\end{tabular}
\raggedright \\
$^a$Total observation time \\
$^b$Mean event rate weighted by the number of pairs for all 10 partial periods \\
$^c$The best-fit value, 68\% CL lower and upper limits of the parameters 
$C$, $p$, and $\tau$ (in [s]) \\
$^d$The branching ratio with the integration upper bound of $\Delta t_u$ where
$\xi(\Delta t_u) = 1$ (top) or 0.1 (bottom) \\
$^e$Before and after the Tohoku earthquake on March 11, 2011.
\label{table:EQ}
\end{table*}

\begin{figure*}
\includegraphics[width=10cm,angle=+90]{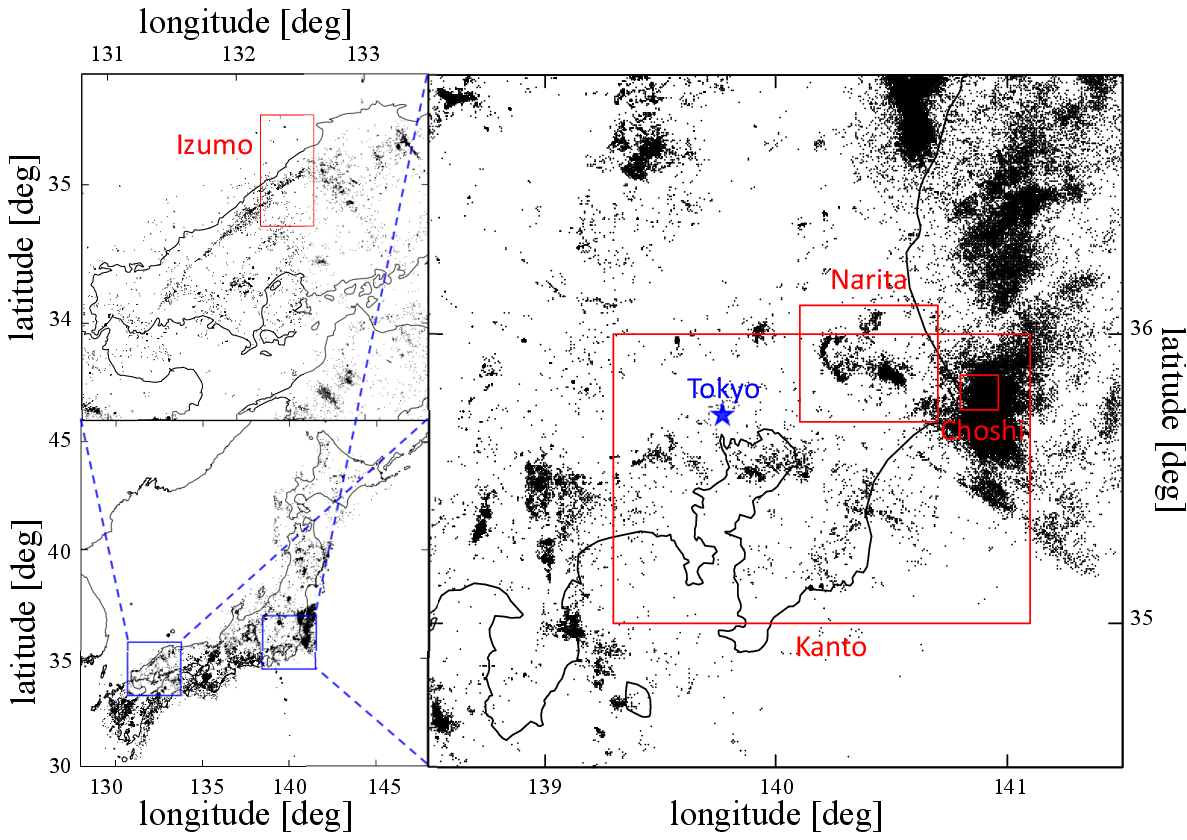}
\caption{Map of regions from which the earthquake data were extracted. 
Distributions of epicenters during a period (May 6, 2010 -- December 31, 2012)
are shown by dots. }
\label{fig:EQ_map}
\end{figure*}

To quantify the similarity to earthquakes, we used the same analysis method to examine 
time-energy correlations of earthquakes. 
The earthquake data used in this study were extracted from the 
Japan Unified hI-resolution relocated Catalog for Earthquakes (JUICE, \citealt{Yano2017}). 
Five data sets of similar sample sizes to FRBs ($\sim$ 1,000 events for each data set)
were extracted from various regions, area sizes, and time periods (Table \ref{table:EQ}). 
The three regions of Choshi, Narita, and Izumo were selected
to see the regional dependences, and data from the greater Kanto region 
(including Choshi and Narita) were also analyzed to see the dependence on area size 
(Fig. \ref{fig:EQ_map}).
To keep the sample size similar to that of FRBs, the time period was adjusted so that 
the number of events in each data set is approximately 1,000.
The rate of earthquakes changed dramatically before and after the huge Tohoku
earthquake on March 11, 2011. We analyzed data from Narita before and after this
to see the impact of the change in activity.
The other three data sets are all before the earthquake.

\begin{figure*}
\includegraphics[width=12cm,angle=-90]{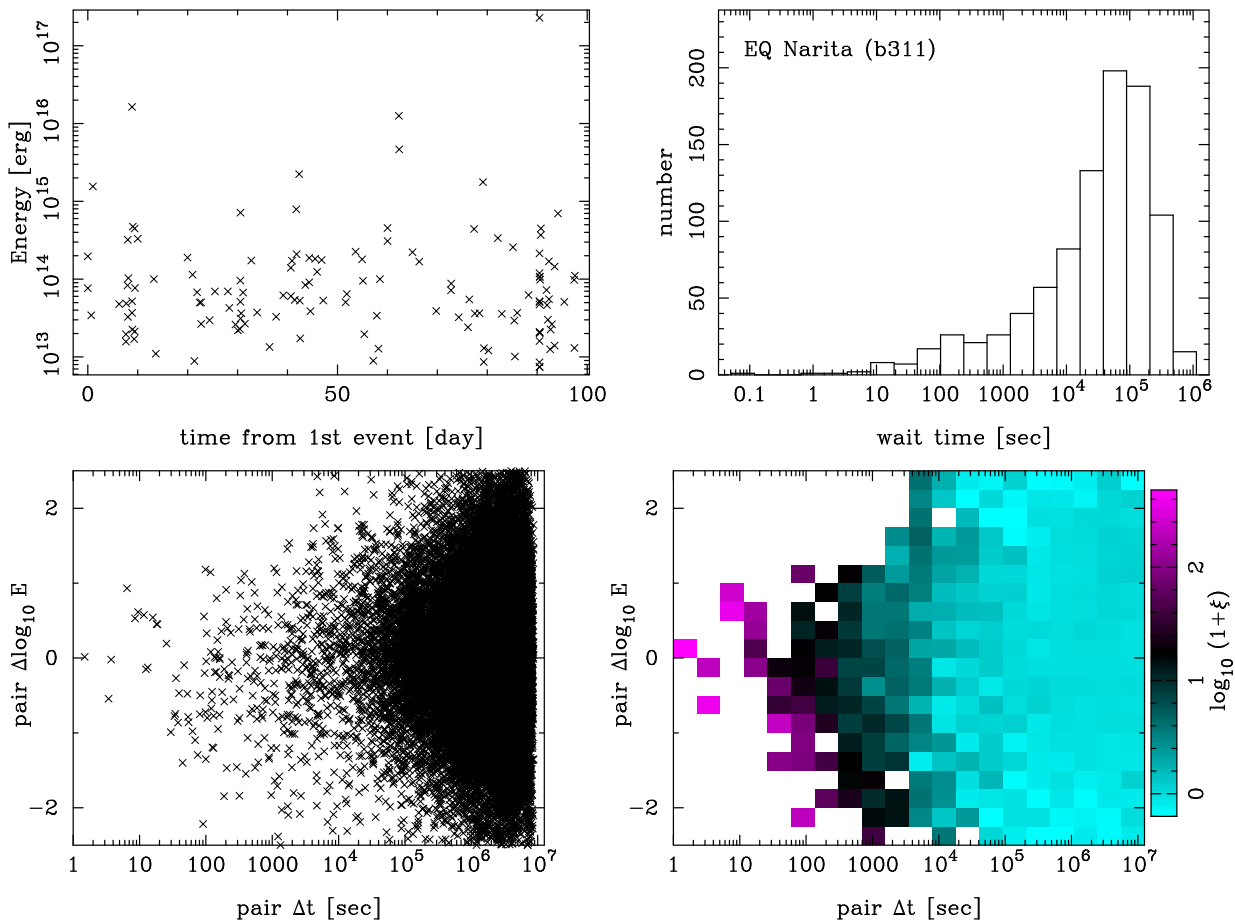}
\caption{ The same as Fig. \ref{fig:FRB_L21}, but for the Narita (before 311) earthquake data. }
\label{fig:EQ_narita_b311}
\end{figure*}

The energy of earthquakes was calculated from the magnitude by the formula
$E = 10^{11.8 + 1.5 M}$ erg. Magnitudes in the catalog are discretized with an 
accuracy of 0.1, and they are shifted randomly with an amplitude of less than $\pm$0.05 
to avoid creating pairs with exactly zero energy differences.
Unlike FRBs, observations of earthquakes and solar flares were made continuously 24 hours a 
day. However, to make the analysis as similar as possible to FRBs, we divided a data set 
into 10 equal partial periods, with each corresponding to a single date of
FRB observations. As with FRBs, pairs were looked for only within partial periods, 
and pair counts were added up over the entire period to compute $\xi$.
Results for the Narita b311 data are shown in Fig. \ref{fig:EQ_narita_b311},
and those of the other four sets are in Figs. \ref{fig:EQ_narita_a311}-\ref{fig:EQ_izumo}
in Appendix \ref{app:figs}. 

\subsection{FRBs versus earthquakes}

\begin{figure}
\includegraphics[width=13cm,angle=-90]{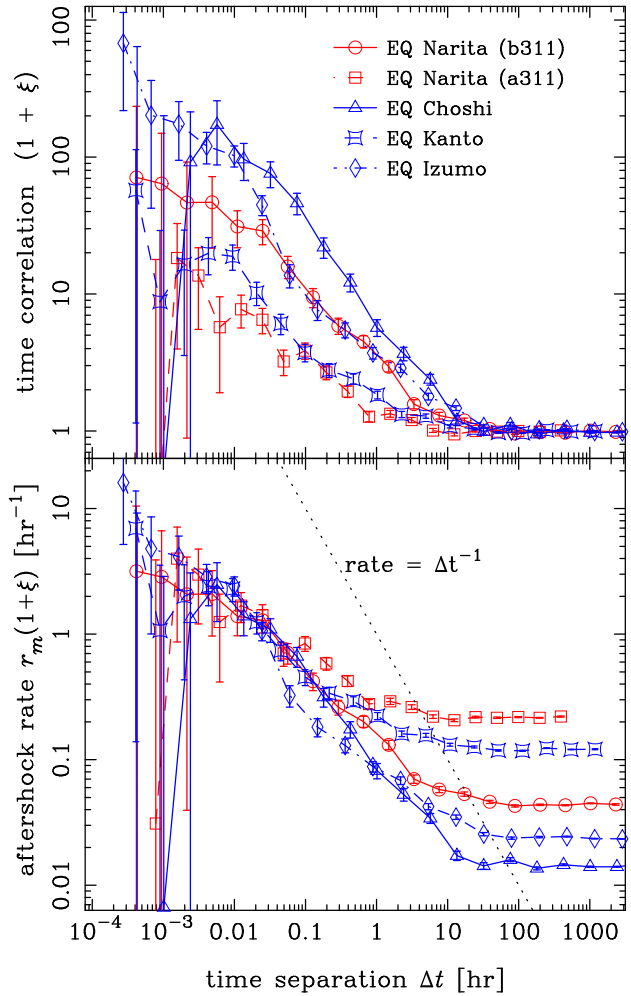}
\caption{ The same as Fig. \ref{fig:FRB_xi1D}, but for the earthquake data.  }
\label{fig:EQ_xi1D}
\end{figure}

\begin{figure*}
\includegraphics[width=8cm,angle=+90]{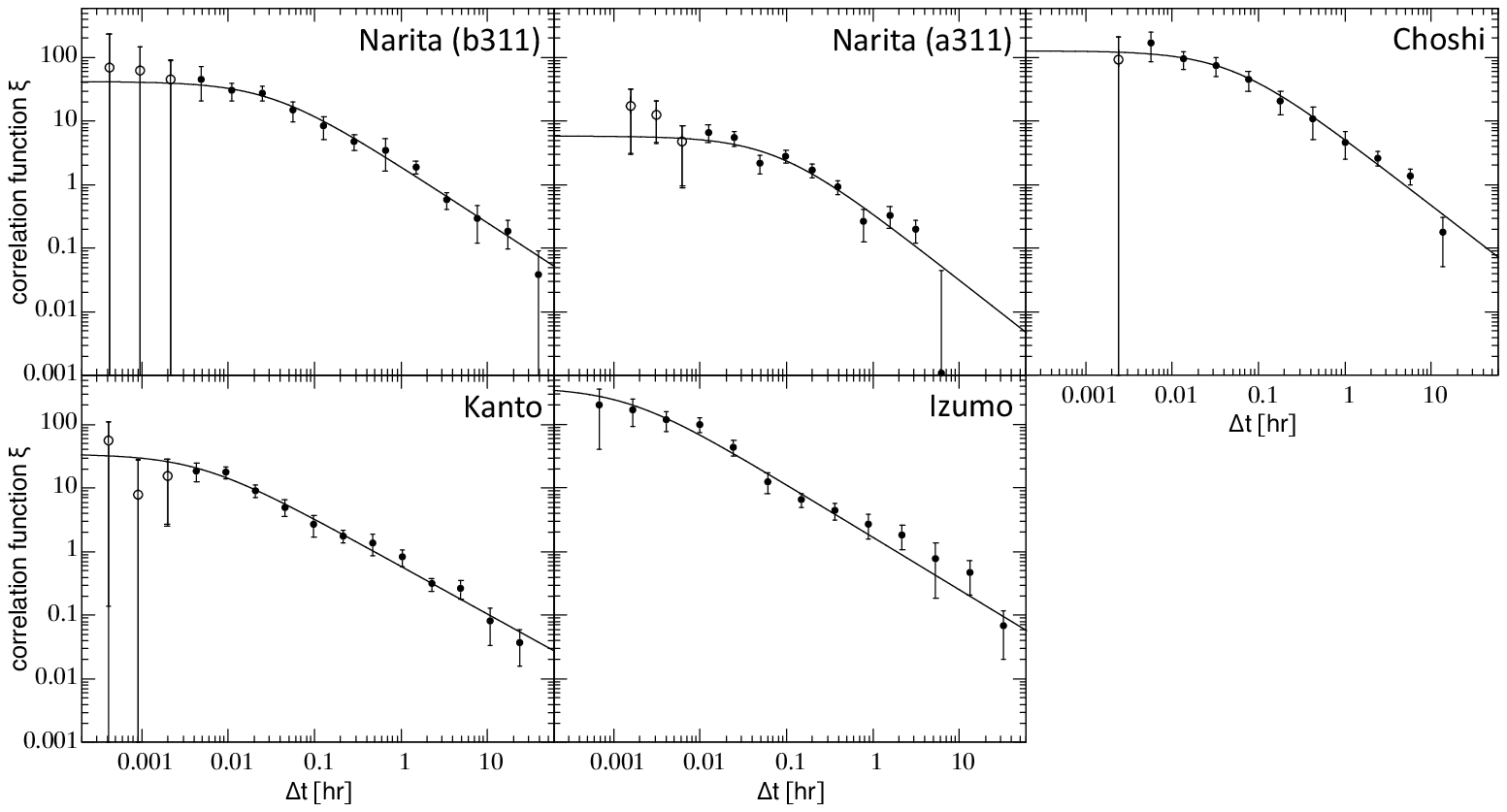}
\caption{ The same as Fig. \ref{fig:FRB_fits}, but for the earthquake data. }
\label{fig:EQ_fits}
\end{figure*}

The time correlation functions of the earthquake data sets are shown
in Figs. \ref{fig:EQ_xi1D} and \ref{fig:EQ_fits}. They are similar to
FRBs in that $\xi(\Delta t)$ is a power law 
of $\propto (\Delta t + \tau)^{-p}$ at small $\Delta t$ but uncorrelated
events at a constant rate become dominant in the large $\Delta t$ region.
Denoting $r_m$ as the mean event rate including uncorrelated events
(shown in Table \ref{table:FRB} and Table \ref{table:EQ}), 
$(1+\xi) \, r_m$ becomes the aftershock rate, i.e., 
the rate of events occurring $\Delta t$ after an event, 
by definition of the correlation function. This is shown
in the bottom panel of Figs. \ref{fig:FRB_xi1D} (FRBs) and \ref{fig:EQ_xi1D} (earthquakes). 
Since FRB event rates vary on different observation dates within 
a single data set, we calculated $r_m$ for the entire data set by taking an average 
weighted by the number of pairs on each date (or partial period in earthquake data). 

The expected number of correlated aftershocks to an event
(called ``the branching ratio'' in earthquake studies) is given by
$n \equiv \int r_m \, \xi(\Delta t) \, d(\Delta t) $, and if $n$ is less than 1, 
the effect of a single event causing multiple aftershocks is small. 
The condition for $n < 1$ is roughly $r_m \, \xi < (\Delta t)^{-1}$, 
which is satisfied in both the FRB and earthquake data, as seen in 
Figs. \ref{fig:FRB_xi1D} and \ref{fig:EQ_xi1D}.
The branching ratio $n$ was calculated using the best-fit parameter values 
of $\xi(\Delta t)$, and shown in Tables \ref{table:FRB} and \ref{table:EQ}. 
Since $p$ is smaller than 1 in some of 
the earthquake data sets, the upper bound $\Delta t_u$ of the integration for $n$
was set to the point where $\xi(\Delta t_u)=1$ or 0.1 to avoid divergence.
The values of $n$ are indeed less than 1, but of order unity ($\gtrsim$ 0.1)
 for all FRB and earthquake data sets.

Then these results can be interpreted as each event produces 
at most one aftershock at a rate consistent with the Omori-Utsu law,
$r_m \, \xi \propto (\Delta t + \tau)^{-p}$, where all events are treated on the same footing
and there is no distinction between mainshocks and aftershocks.
Such a model is called ETAS (epidemic-type aftershock sequence)
in earthquake studies and is known to explain earthquake data 
well \citep{Ogata1999,Saichev2006,deArcangelis2016}. 

Further similarities between FRBs and earthquakes can be pointed out. 
The branching ratio $n$ is about 0.1--0.5 for both FRBs and earthquakes 
(Tables \ref{table:FRB} and \ref{table:EQ}), which is
also similar to those found in past earthquake studies \citep{Saichev2006,deArcangelis2016}.
The power law flattens at $\Delta t \lesssim \tau$, and $\tau$ 
is about 1--10 msec for FRBs and 0.3--3 min for earthquakes, which are close to the typical 
duration of the respective phenomena. It should also be noted that 
the correlated aftershock rate ($r_m \, \xi$) is stable both in FRBs  and earthquakes 
(bottom panels of Figs. \ref{fig:FRB_xi1D} and \ref{fig:EQ_xi1D}), even though 
the mean event rate $r_m$ of uncorrelated events varies widely with changes of
source activity (and also with the choice of region and its area size for earthquakes).
The aftershock rates of the three different FRB sources are not significantly different, 
but the rate fluctuations are even smaller within each source.
This indicates that aftershocks are not caused by the activity of the source as a whole, 
but by the changes induced by each individual event.
Finally, there is little correlation between $\Delta t$ and $\Delta \lg E$ in either FRBs or
earthquakes ($\xi$ almost constant against $\Delta \lg E$ in
Figs. \ref{fig:FRB_L21},  \ref{fig:EQ_narita_b311}, and
\ref{fig:FRB_H22}--\ref{fig:EQ_izumo}).

In fact, a slight time-magnitude correlation has been reported by detailed
statistical tests for earthquakes \citep{Lippiello2008,deArcangelis2016,Zhang2023b}. 
Asymmetry with respect to $\Delta \lg E$ (more negative $\Delta \lg E$ pairs, i.e., 
aftershock energy smaller than mainshock)
may also be evident in some of our earthquake data sets:
Narita (b311, Fig. \ref{fig:EQ_narita_b311}), Choshi (Fig. \ref{fig:EQ_choshi}),
and Izumo (Fig. \ref{fig:EQ_izumo}). 
Similar asymmetry about $\Delta \lg E$ is not visible in the FRB data,
but it may be due to a narrow dynamic range of $E$ caused by detection limits.
Careful verification by future studies would be desirable.

The only difference between the two phenomena
is the value of the Omori-Utsu index, $p$:
FRBs have a larger $p$ resulting in the bimodal wait-time distribution, 
while bimodality is not clearly visible in the earthquake data. 
It should be noted, however, that $p$ of earthquake data varies widely 
depending on regions, and those of FRBs are marginally within the range 
for earthquakes ($p = $ 0.6--1.9) reported 
by past studies \citep{Utsu1995,Ogata1999,Wiemer1999,deArcangelis2016}.

\subsubsection{Notes on previous studies on the Omori-Utsu law in
neutron star phenomena}

\cite{Wang2018} found that the rate evolution of 14 bursts detected during an observation
of about 60 minutes from FRB 20121102A is consistent with the Omori-Utsu law. 
This result is likely unrelated to the correlations found in this study because the 
correlation time scale 
of \cite{Wang2018} is much larger than $\Delta t \sim$ 1 s, where we did not find significant 
correlations. It is difficult to draw strong conclusions from the sample of \cite{Wang2018},
not only because of the small statistics but also because of
the arbitrariness of where to separate mainshocks and aftershocks.
It is possible that the rate variation observed by \cite{Wang2018}
does not reflect a systematic aftershock sequence, but rather 
a change in the source activity during that observation period.

\cite{Enoto2017} analyzed the light curves of X-ray outbursts in three magnetars and found 
that they can be fitted by the Omori-Utsu law, and proposed the possibility that X-ray luminosity
is superpositions of many small starquakes. 
However, the plateau time scale $\tau$ is longer than 10 days, which is many orders of magnitude
larger than the time correlations found in this work, suggesting that there is no direct relationship
between the two. It should be noted that continuous luminosity evolution can also be explained by
continuous evolution of energy release in a single system, 
rather than by the superposition of many discrete aftershocks. 
In general, in systems where the energy loss rate $\dot E$ is proportional to the power
of the system energy $E$ ($\dot E \propto E^q$), the evolution of $\dot E$ takes the
form of the Omori-Utsu formula with $p = q/(q-1)$ (e.g., $q=2$ for a rotation-powered pulsar
with a constant magnetic field strength).

\subsection{Comparison with solar flares}

\begin{figure}
\includegraphics[width=13cm,angle=-90]{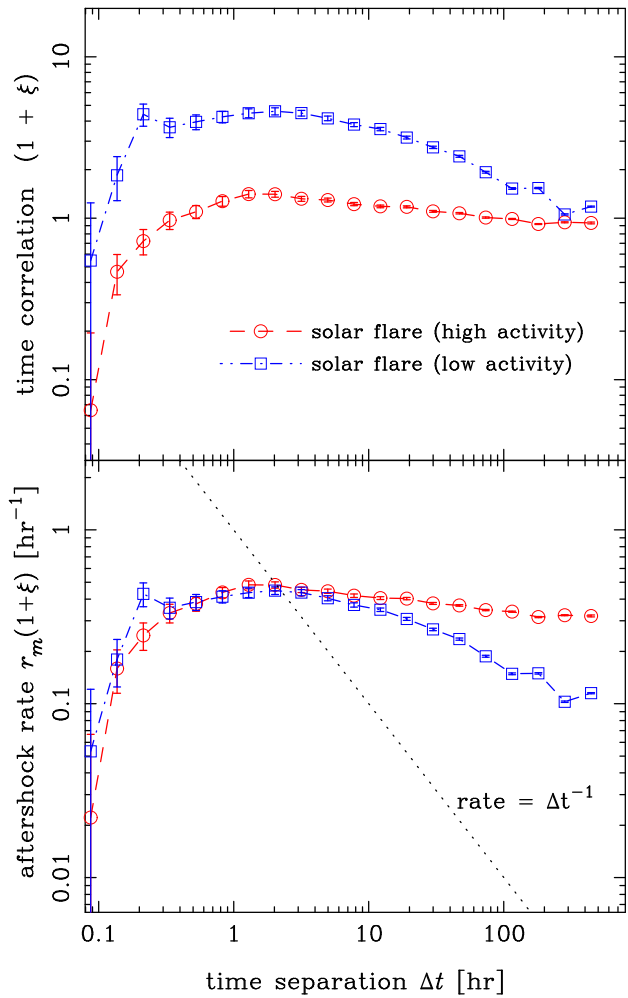}
\caption{ The same as Fig. \ref{fig:FRB_xi1D}, but for the solar flare data. }
\label{fig:SF_xi1D}
\end{figure}

The solar flare data were taken from the Hinode catalog \citep{Watanabe2012}.
To see the dependence on activity level, data were extracted from two periods 
(200 days from 2012 Apr. to Oct., and 1,200 days from 2017 Oct. to 2021 Jan.) of high and low
solar activity, adjusting the time period so that each period includes about 1,000 events
(1,422 and 1,207 events, respectively).
The X-ray flux was calculated from the flare class in the catalog, and this was used as a proxy 
for energy $E$. Flare classes are given as two significant digits and, as with the earthquake 
data, they were randomly shifted by the precision of the significant digits
[e.g., X-ray flux of a B4.2 class flare is randomly chosen from 
the range of $(4.15$--$4.25) \times 10^{-4} \ \rm erg \ cm^{-2} \ s^{-1}$].
These data sets were analyzed in the same way as earthquakes, 
and the results for the low activity set is shown in Fig. \ref{fig:SF_low}
(and the high activity set in Fig. \ref{fig:SF_high} in Appendix \ref{app:figs}). 
Time correlation functions for the two sets are shown in Fig. \ref{fig:SF_xi1D}. 

\begin{figure*}
\includegraphics[width=12cm,angle=-90]{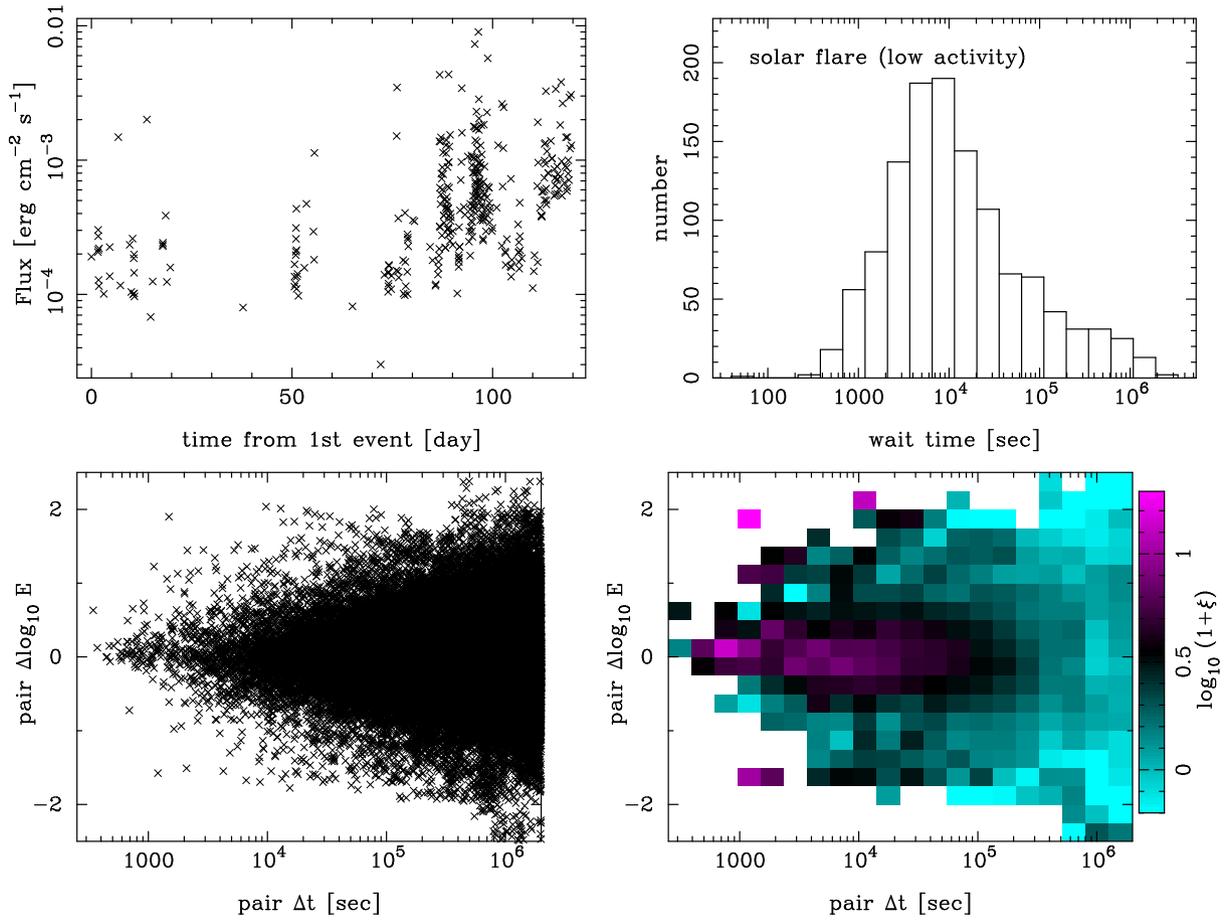}
\caption{ The same as Fig. \ref{fig:FRB_L21}, but for the solar flare (low activity) data. }
\label{fig:SF_low}
\end{figure*}

Figure \ref{fig:SF_xi1D} shows that a significant time correlation signal is detected 
in the region of $\Delta t \lesssim 100$ hr. 
It is known that flares occurring in different active regions on the solar surface are uncorrelated 
\citep{Benz2017}, and hence the correlated signals are thought to be between flares occurring 
in a common active region.
We can convert $\xi$ into the correlated aftershock rate $r_m \, \xi$, 
and it is stable regardless of the 
level of solar activity ($r_m$), which is similar to the behaviors of FRBs and earthquakes. 
However, $\xi$ reaches at most 5 without showing a clear power law, in sharp contrast to FRBs 
and earthquakes. The correlated rate $r_m \, \xi$
is well above $(\Delta t)^{-1}$, which indicates that the branching ratio $n$ is larger than 1
and multiple aftershocks are generated from a single event. Another difference from FRBs or
earthquakes is the strong correlation not only in time but also in energy,
with $\xi$ peaking at $\Delta t \sim 10^3$--$10^4$ s and $\Delta \lg E \sim 0$
(lower-right panels of Figs. \ref{fig:SF_low} and \ref{fig:SF_high}). 
This strong time-energy clustering nature is evident 
in the $t$-$E$ distribution shown in these figures (upper-left panels),
which are not seen in FRB or earthquake data. 

Though similarities between solar flares and magnetars are often discussed,
these results indicate a different nature of the time-energy correlation of solar flares
compared to FRBs and earthquakes. 
Interpretation of these results in comparison with the physics and theory of solar flares is 
outside the scope of this study. 
Time and/or energy correlations of solar flares have been discussed by analyzing 
the wait time distributions and cross-correlations between energy and
time (either since the last event or to the next event) 
\citep{Wheatland2000,Lippiello2010,Hudson2020,Carlin2023}.
It is not straightforward to compare our results with these, because $\Delta t$ 
in our analysis is not wait time but time separation for arbitrary pairs, and 
we are looking at the energy difference $\Delta \lg E$
of an event pair, rather than the energy itself of one event. 
Careful comparison of these results to elicit information about the physics of solar flares is an
interesting subject for future research. In such studies,
it is important to take into account observational biases in the solar flare catalog, 
such as the obscuration effect (lower detection efficiency for faint flares after a giant flare,
\citealt{Wheatland2001}).

\section{Conclusions and Discussion}

By examining the correlation functions in time-energy space, we found remarkable similarities
between the statistical properties of FRBs and earthquakes, especially the laws on aftershock
occurrence. Listing the similarities, (1) the aftershock rate
follows a power law of $\propto (\Delta t + \tau)^{-p}$ (the Omori-Utsu law), (2) $\tau$ 
is about the same as the typical event duration, (3) the branching ratio
(expected number of aftershocks associated with a single event) is $n = 0.1$--0.6 
for both phenomena, (4)  the correlated
aftershock rates remain stable regardless of the change in the source activity or mean
event rate, and (5) there is little correlation between energy and time. 

In contrast, the correlation functions for solar flares differ significantly from those of FRBs 
and earthquakes. The correlation function cannot be fitted by a power law, and the branching ratio
is significantly greater than unity, which means that a single event causes multiple aftershocks. 
There is also a strong correlation in the direction of energy, with flares of similar energy tending 
to occur in succession. It is well known that
the energy distribution of solar flares follows a power law,  which is often noted to be
similar to that of earthquakes (the 
Wadati-Gutenberg-Richter law\footnote{This is often referred to as the Gutenberg-Richter law, but
see \citet{Utsu1999,deArcangelis2016} for a historical account.},
\citealt{Wadati1932,Gutenberg1944}). 
However, with respect to correlations in the time-energy space,
solar flares do not resemble earthquakes, and by comparison, the similarity between 
FRBs and earthquakes is remarkable. 

Both solar flares and magnetars are caused by magnetic energy, but unlike the Sun, 
neutron stars are thought to have solid crusts on their surfaces.
Therefore, the most natural interpretation of the present results is that FRBs are 
earthquake-like phenomena that suddenly release the energy stored in neutron star crusts. 
Other models of FRB repeaters are not immediately dismissed, but any model must be 
consistent with this time-energy correlation, placing strong constraints on possible models.

In the case of earthquakes, differences in the index $p$ are thought to reflect the physical
properties of crust and seismic 
processes \citep{Ogata1999,Wiemer1999,deArcangelis2016}. 
Then relatively large $p$ values of FRBs may provide information about the physical properties
of neutron star crusts, and energy production mechanism by seismic processes in them. 
This suggests a new possibility for future studies to probe the physics
of dense nuclear matter by using repeater FRBs.

Two FRB repeaters, including 20121102A, are known to periodically
change their activity in cycles of 16 or 160 
days \citep{CHIME2020-period,Rajwade2020,Cruces2021}. 
If these are neutron stars in a binary system, the activity may be increased when
the crust is deformed by tidal forces from the companion star at the pericenter. 
Models of binary origin have been proposed for the periodic FRB activities
\citep{Ioka2020,Lyutikov2020,Barkov2022}, but most have 
assumed that the periodicity is due to the absorption of FRB radiation 
(e.g. by stellar winds from the companion star), rather than intrinsic change
of the FRB production activity. 
A repeater FRB has been found 
in a globular cluster of the nearby galaxy M81 \citep{Kirsten2022}, 
and the FRB activity may be stimulated by close encounters with other stars in the cluster.

As far as we know,  most previous studies 
on the time correlation of earthquakes or solar flares
have also been based on 
wait times rather than correlation function \citep{deArcangelis2016,Saichev2006,Wheatland2000}.
The correlation function method adopted here
fully exploits the temporal information of the sample, 
and the time dependence of the correlated aftershock rate can be seen more
directly. This is why the two phenomena (FRBs and
earthquakes), that appear to be different in the 
wait time distribution, turn out to have essentially the same properties when viewed 
in terms of the correlation function. Applying this method 
to a wider range of earthquake and solar flare data may provide new insights 
into these phenomena.

\section*{Acknowledgements}

We thank Pei Wang, Di Li, Dant\'e Hewitt, Heng Xu, Joscha Jahns, and Yongkun Zhang
for providing the full numerical data of their FRB catalog and/or related information. 
TT was supported by the JSPS/MEXT KAKENHI Grant Number 18K03692.

\section*{Data Availability}
 
The data newly derived in this article (e.g., correlation function values) 
will be shared on reasonable request to the corresponding author.




\bibliographystyle{mnras}
\bibliography{totani2023} 




\appendix

\section{Figures of detailed results for the data sets not presented in the main part}
\label{app:figs}

\begin{figure*}
\includegraphics[width=11cm,angle=-90]{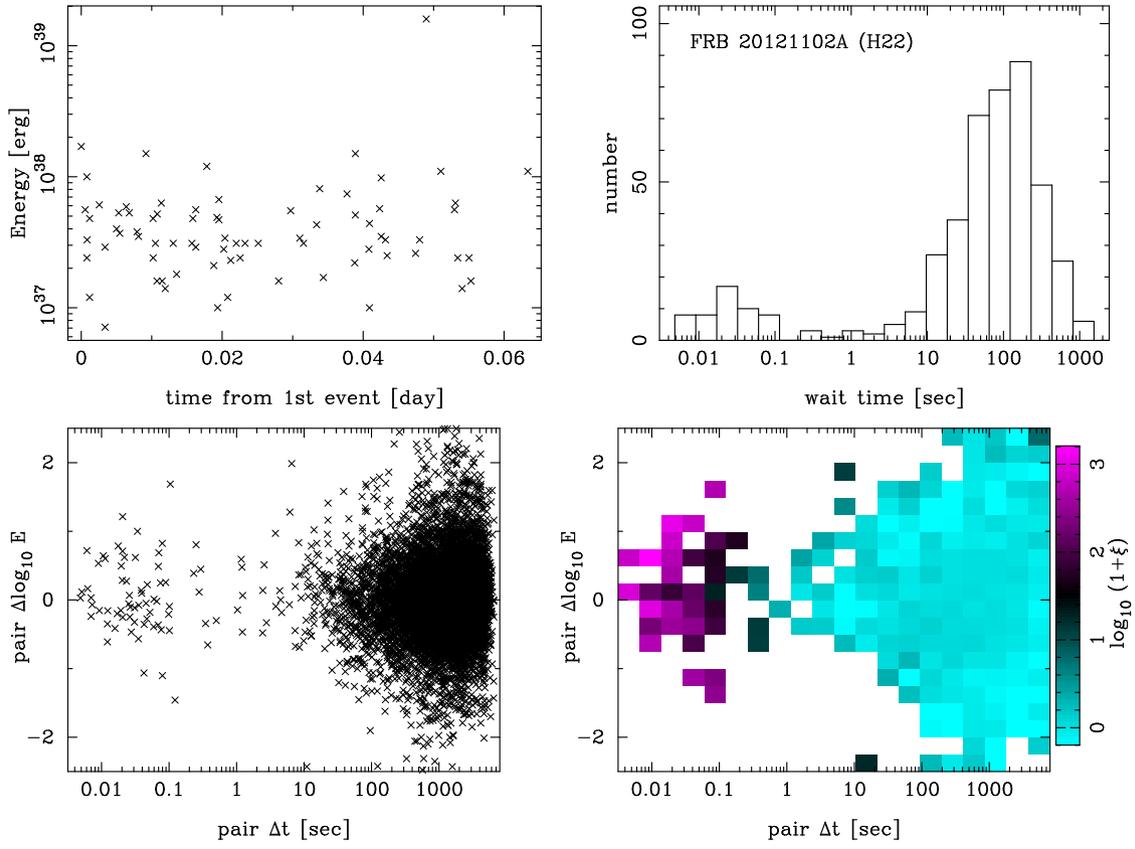}
\caption{ The same as Fig. \ref{fig:FRB_L21}, but for the FRB 20121102A (H22) data. }
\label{fig:FRB_H22}
\end{figure*}

\begin{figure*}
\includegraphics[width=11cm,angle=-90]{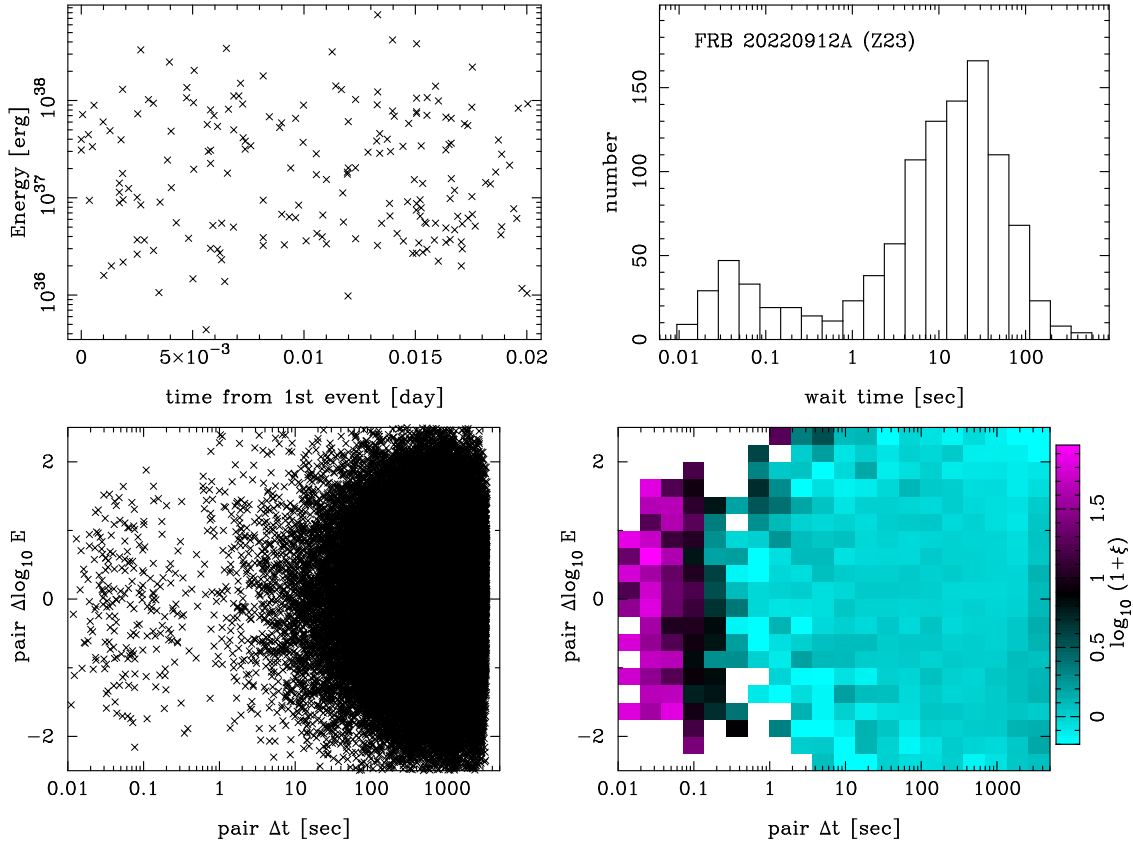}
\caption{ The same as Fig. \ref{fig:FRB_L21}, but for the FRB 20220912A (Z23) data.  }
\label{fig:FRB_Z23}
\end{figure*}

Figures of detailed results ($t$-$E$ scatter plot, wait-time distribution, pair distribution
and correlation function in $\Delta t$-$\Delta \lg E$ space) for each data set
were presented
only for FRB 20121102A (L21, Fig. \ref{fig:FRB_L21}), earthquake Narita (b311, Fig. 
\ref{fig:EQ_narita_b311}), and solar flare (low activity, Fig. \ref{fig:SF_low}) 
in the main part. Here figures for other data sets are presented
as Figs. \ref{fig:FRB_H22}--\ref{fig:SF_high}. 

\begin{figure*}
\includegraphics[width=11cm,angle=-90]{./figs/FRB121102_J23_wsub.ps}
\caption{ The same as Fig. \ref{fig:FRB_L21}, but for the FRB 20121102A (J23) data.  }
\label{fig:FRB_D23}
\end{figure*}

\begin{figure*}
\includegraphics[width=11cm,angle=-90]{./figs/FRB201124_X22.ps}
\caption{ The same as Fig. \ref{fig:FRB_L21}, but for the FRB 20201124A (X22) data.  }
\label{fig:FRB_X22}
\end{figure*}

\begin{figure*}
\includegraphics[width=11cm,angle=-90]{./figs/FRB201124_Z22_D3.ps}
\caption{ The same as Fig. \ref{fig:FRB_L21}, but for the FRB 20201124A (Z22, D3) data.  }
\label{fig:FRB_Z22_D3}
\end{figure*}

\begin{figure*}
\includegraphics[width=11cm,angle=-90]{./figs/FRB201124_Z22_D4.ps}
\caption{ The same as Fig. \ref{fig:FRB_L21}, but for the FRB 20201124A (Z22, D4) data.  }
\label{fig:FRB_Z22_D4}
\end{figure*}

\begin{figure*}
\includegraphics[width=11cm,angle=-90]{./figs/EQ_narita_a311.ps}
\caption{ The same as Fig. \ref{fig:FRB_L21}, but for the Narita (after 311) earthquake data. }
\label{fig:EQ_narita_a311}
\end{figure*}

\begin{figure*}
\includegraphics[width=11cm,angle=-90]{./figs/EQ_choshi.ps}
\caption{ The same as Fig. \ref{fig:FRB_L21}, but for the Choshi earthquake data. }
\label{fig:EQ_choshi}
\end{figure*}

\begin{figure*}
\includegraphics[width=11cm,angle=-90]{./figs/EQ_kanto.ps}
\caption{ The same as Fig. \ref{fig:FRB_L21}, but for the Kanto earthquake data. }
\label{fig:EQ_kanto}
\end{figure*}

\begin{figure*}
\includegraphics[width=11cm,angle=-90]{./figs/EQ_izumo.ps}
\caption{ The same as Fig. \ref{fig:FRB_L21}, but for the Izumo earthquake data. }
\label{fig:EQ_izumo}
\end{figure*}

\begin{figure*}
\includegraphics[width=11cm,angle=-90]{./figs/SF_high.ps}
\caption{ The same as Fig. \ref{fig:FRB_L21}, but for the solar flare (high activity) data. }
\label{fig:SF_high}
\end{figure*}


\bsp	
\label{lastpage}
\end{document}